# A Labeling Task Design for Supporting Algorithmic Needs: Facilitating Worker Diversity and Reducing AI Bias

Jaeyoun You[1], Daemin Park[2], Joo-yeong Song[3], Bongwon Suh[1]

[1] *Graduate School of Convergence Science and Technology, Seoul National University, Seoul, Republic of Korea*

[2] *Sunmoon University, Asan, Republic of Korea*

[3] *Obzen, Seoul, Republic of Korea*

Corresponding author: Bongwon Suh, bongwon@snu.ac.kr

# A Labeling Task Design for Supporting Algorithmic Needs: Facilitating Worker Diversity and Reducing AI Bias

Studies on supervised machine learning (ML) recommend involving workers from various backgrounds in training dataset labeling to reduce algorithmic bias. Moreover, sophisticated tasks for categorizing objects in images are necessary to improve ML performance, further complicating micro-tasks. This study aims to develop a task design incorporating the fair participation of people, regardless of their specific backgrounds or task's difficulty. By collaborating with 75 labelers from diverse backgrounds for 3 months, we analyzed workers' log-data and relevant narratives to identify the task's hurdles and helpers. The findings revealed that workers' decision-making tendencies varied depending on their backgrounds. We found that the community that positively helps workers and the machine's feedback perceived by workers could make people easily engaged in works. Hence, ML's bias could be expectedly mitigated. Based on these findings, we suggest an extended human-in-the-loop approach that connects labelers, machines, and communities rather than isolating individual workers.

## 1 INTRODUCTION

Image annotations with labeled features made by humans can be utilized as a training dataset for artificial intelligence (AI) (Deng et al., 2009; Krizhevsky et al., 2012; Simonyan & Zisserman, 2014). Despite the algorithm's high performance, however, some programs, such as face recognition systems and job recommendation programs trained by human data, reveal racial and sexual biases (Buolamwini & Gebru, 2018). It has been demonstrated previously that AI systems fail to interpret inputs from people with disabilities, and adapting AI systems to realize efficient interaction with people with disabilities is a necessity (Guo et al., 2019). Therefore, researchers attempt to provide evidence about potential biases brought by a dataset constructed by labelers with skewed demographic backgrounds, pointing out that human decisions vary in different cultures (Awad et al., 2018).

With the gradual improvement of AI, labeling tasks also become more complex and challenging. Thus far, many labeling tasks were often uncomplicated and easy to do for human labelers, regardless of their specific backgrounds (e.g., separating human faces with dogs in a video clip was a trivial task for all people) (Yuen et al., 2011). However, to enhance the algorithm's performance, we need to use more specifically categorized labels for more specific object detection. For instance, images classified as Maltese, Pomeranian, German Spitz, among others, could be more informative and useful than images annotated simply as "dog."

A more subjective decision-making process leads to more hurdles for people to participate in tasks because 1) to choose multiple objects, whatever is perceived in the video possibly depends on individuals; and 2) unless categories are properly predefined and easily memorized, human labelers generate results with individually biased decisions among multiple selections. It is stated that many of the labelers may have a problem in selecting the label in the image within many label categories, leading to a highly uncertain decision making.

Even if the task becomes more comfortable and lowers the hurdles for all, it is not easy to balance labelers' ratio by demographic criteria. Although subjective bias is a crucial issue, it is not still sufficiently considered in image annotation tasks due to strong arguments supporting collective intelligence's effectiveness – a large number of opinions would end up with proper decisions (Bountouridis et al., 2019; Ye & Robert, 2017). However, methods have been proposed to extract meaningful inputs from different answers from individuals, and meaningful responses have been extracted from data previously considered noise (Kairam & Heer, 2016). Maintaining variety and balance among training data are a desirable property.

In one notable study, researchers assert strong interventionists to the data collection process to achieve fair participation and less subjective biases (Jo & Gebru, 2020). They propose solutions derived from archival practices. Such as a collection policy or mission statements of data collection should be applied to the ML data collecting process to mitigate biased labels. The paper indicates the probability to socio-cultural bias in text database and claimed to make guidelines opposite to the laisser-faire circumstance of ML data collection. Our theoretical basement also lay in need of intervention in the data collection process.

As appropriate intervention is required in a complex labeling work environment, we would like to address the following research questions.

- **RQ 1.** How does bias appear when label complexity and various demographic configurations are mixed?
- **RQ 2.** Are there any invisible influencing factors in the current complex labeling process?

In this paper, we begin with checking biased patterns among in the labeling micro-task. Detecting biased pattern among demographical groups offers a strong hypothesis to the study to identify the main problems existing in complex labeling tasks. We then find essential features that help and encourage workers in tasks by analyzing the labeler's narrative data from a community chat for 3 months. Based on the findings, we designed the suggestion model – the extended human-in-the-loop (eHITL) – to addressing a better micro-task process. The model describes the loop connecting the human labelers to the community that helps workers, and the machine provides human psychological rewards, described in Fig. 1. We suggest this task model design, which could guarantee fair participation in micro tasks and increase the dataset's diversity, thereby mitigating algorithmic biases. While the previous HITL model included only two stakeholders, the individual person and the machine, the eHITL model includes another group of individuals as a stakeholder who enthusiastically assists individual crowdworkers.

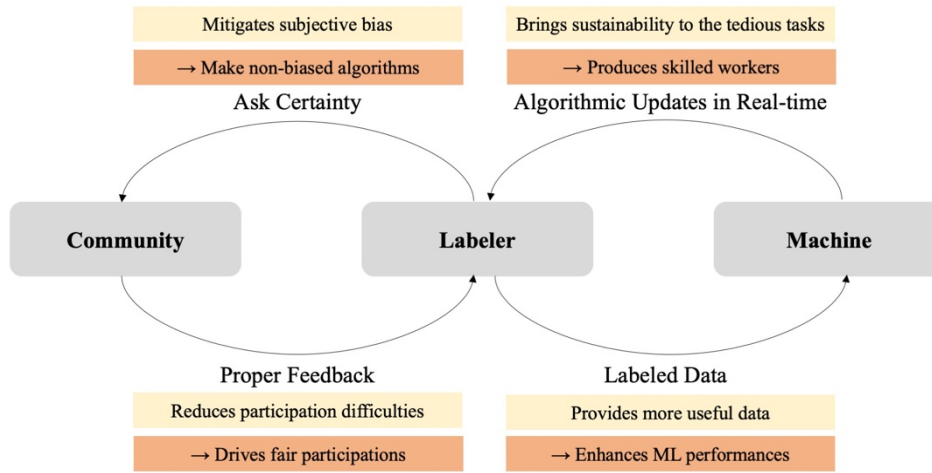

Fig. 1. The proposed model: extended human-in-the-loop (eHITL) model in the micro-task design.

This study is based on a **large-scale fieldwork scenario**. We observed the three-month labeling process with participants, tools and annotation manuals. Using the data collected in this study, we determine if any **personal bias** exists **in the output labels**. Then, we investigate the possibility of realizing **fair participation and mitigating subjective biases**. Based on the results of the qualitative and quantitative analyses described Section 4, we finally propose the task design model – *eHITL in Section 5*. In addition, we discuss eHITL in the context of micro-task projects in the Discussion section.

**2 RELATED WORK**

This section reviews previous studies that provided us with insights into micro-task design. In reviewing previous works related to our research questions, we considered two main topics: the participation of people from diverse backgrounds and the effect of tools they use in their crowdsourcing tasks.

*2.1 Accessibility for Micro-task Workers*

In early studies related to micro-workers' behavior, the efficiency of crowd workers was mainly studied as follows. Kittur et al. (2008a; 2008b) thoroughly investigated crowds who produce better outputs in early crowd work studies. Their study was extended to numerous collaborative work studies, such as MathOverflow (Tausczik et al., 2014) and real-time collaboration systems (Lee et al., 2018). Moreover, strategies to manage anonymous workers have been studied from multiple perspectives. Previous studies have proposed free-rides prevention designs (Zytko et al., 2018), studied better rewards for workers (d'Eon et al., 2019; Yu et al., 2014), and offered certifications and developed stellar reputations as rewards for individual labelers (Kittur et al., 2013).

Recent studies have focused more on individual workers, whereas past studies have addressed effective crowdsourcing work platform designs. For studies on individual worker motivation, Kaufmann et al. (2011) suggest categories with intrinsic and extrinsic factors to analyse. Specifically, Kobayashi et al. (2015) explored elderly workers in socially motivated crowdsourcing, and Brewer et al. (2016) investigated the awareness of elderly workers regarding crowd work more in detail. They suggests that understanding the purpose of crowd work and overcoming accessibility issues would encourage the participation of elderly workers. Further, Chiang et al. (2018) proposed to posit a peer coaching system to support workers' skill development. Their proposed idea has the possibility to maintain worker's sustainability to micro-tasks and to improve data quality.

The accessibility to crowd work has garnered increasing attention. Swanminathan et al. (2017) demonstrated that Amazon's mechanical turk (MTurk) has poor accessibility for people with disabilities, and it was suggested that crowd work design technology is required to ensure accessibility. It was also demonstrated that the crowd work interface should be configured to correct accessibility problems automatically. Zyskowski et al. (2015) pointed out the necessity of customized labor platforms for people with disabilities. Ding et al. (2017) specified "socially embedded work" as a crowd work system for wheelchair users, explaining the gatekeeper structure in online work. Their results indicate that environments fostering social relationships help people with disabilities engage in crowd work. Liu et al. (2016) have reported an in-depth interview on people with disabilities who are able to use Internet and technology and want to help others using these platforms. Our approach is also based on the possibility of people with disabilities contributing to society.

Studies about community-based work for crowdsourcing also have been received attention. Wang et al. (2020) suggested the "Crowdfarm" concept to form government-driven crowdsourcing work in China. Researchers explained that China's socially constructed work shows community-based tasks, and not individual work. In this study, we also focused on the presence of communities that help workers, semblable to the Crowdfarm case. *We extended the range of workers in the crowd work ecosystem, which is not limited to people with similar backgrounds*.

*2.2 Tools for Better Labeling*

Annotation tools for labelers also have been studied for a better labeling output. While ambiguous guidelines hinder and degrade work quality, previous studies suggested various tool designs and considerations for the countermeasures. Ahn & Dabbish (2004) introduced gamification on labeling system for encouraging people to engage image labeling tasks. Chang et al. (2017) proposed a model that eliminates the burden of creating detailed label guidelines by incorporating ideas from expert annotation workflows. For estimating workers' certainty in annotating, Chung et al. (2019) suggested using the annotation granularity to the tool. For minimizing any bias in a labeling project, using more than two labeling tools was also suggested (Song et al., 2018; Song et al., 2019). The study claimed that individual tool designs could introduce biases depending on their subjective design considerations. Thus, a multi-tool approach was proposed to aggregate accuracy in semantic image segmentation. However, most of the tools using in fieldwork cases are already made and produced before the project starts. We did not conduct a specific review for the tool design utilized in this study. However, in our model description, we focused on how the malfunctioning tool and ambiguous categories affect labelers discouraged to the task.

The ambiguous tool design derives biased outputs, but the data collection process could also lead to an unfair dataset. Results from individual workers with different backgrounds have also been investigated. Dong et al. (2012) showed that cultural effects can be embedded in data from crowdsourcing works. They analyzed labeling patterns and identified different classes between people from European, American, and Chinese backgrounds. Otterbacher et al. (2018) revealed that sexist people perceive less gender biases in images. Sen et al. (2015) showed that gold standards, which refer to domain specific labels created by humans from ground truth, also differed among cultural communities.

Based on these works, several studies on leveraging the diversity of workers have been reported. For feeding more diverse training data into programs, individual characteristics for crowd works also have been investigated. Kazai et al. (2012) claimed that crowd systems should collect each worker's demographic and personal data to solve micro-tasks' anonymity and assess their work quality. For globally diverse participation in data collection, Chi et al. (2019) asserted that geographically diverse images should be deployed in crowdsourcing tasks. They aimed to collect images from users worldwide to add global diversity to training datasets for algorithms. Venkatagiri et al. (2019) also proposed the image geolocation model based on local users' knowledge. Barbosa and Chen (2019) also proposed a framework for crowdsourcing systems to consider factors such as demography, and they showed that demographic bias can be mitigated, and the contributors' income can be increased. Our work is theoretically based on Barbosa and Chen's work.

The primary conclusion that can be made from the above discussion is that diverse users generate diverse data. We positively support the need for diversity in workers and data. *Therefore, we aimed to design the micro-task participation more open, which guarantees universal participation.*

### 2.3   *Human-in-the-Loop (HITL) Model*

We skipped into the HITL model framework because of the rise of its importance in task procedure design. To realize decision-making processes that interact with machines, the HITL model has been considered a basis for human–AI collaborative work design. HITL was first employed in embodied devices for the human body. Minsky et al. (2013) stated the notion in terms of humanistic intelligence ("cyborg"), where the sensor attached to the body/brain creates a feedback loop for synthesis. Recently, HITL has been utilized to overcome the limitations of various algorithms, such as fairness detections and sociotechnical systems (Green & Chen, 2019; Wolf et al., 2018).

HITL model has been vigorously explored in recent studies. This model, which the human and the machine mutually stimulate, helps in a more enhanced decision-making (De-Arteaga, Fogliato & Chouldechova., 2020). People select features or opinions based on their ground knowledge, and the machine learns their idea immediately. Then, the machine feedback returns their output to be evaluated by human users (Holzinger, 2016). Unlike AI-assisted programs, the HITL model performs more interactive flow with humans. By user's incorporation, the machine could improve performances with explainable transparency. This issue has been applauded for the solution to the black-box problems (Holzinger et al., 2017). Further, HITL could explain the role and the effect of the machine's feedback on humans in terms of embodiment (Frohner et al., 2019). From those aspects, we expected to observe the tool's gradual improvement, which affected micro-workers.

In crowd work studies, researchers also consistently have focused on interactive tools for workers and machines. For instance, Sorokin et al. (2010) suggested a human–machine interaction design concept, wherein the entire working process being interactive with micro-workers. Furthermore, Tubaro et al. (2020) defined humans' role in supporting AI as the trainer, verifier, and imitator by analyzing human involvement in the algorithm's workflow. Based on the previous studies, we strongly considered the diversity of people who participate in those interactive flows.

However, as far as we know, the interactive tool cannot fully explain the loop between human and machine because interactive tools generally cannot spontaneously change themselves by receiving human feedback. Algorithmic feedbacks have previously been reviewed as motivational factors for crowdsourcing works (Yu et al., 2014; Jun et al., 2017), and intrinsic/extrinsic motivations for quality control have been identified to increase worker engagement (Daniel et al., 2018). However, the HITL model has not been exactly mentioned and deployed to micro-tasks yet; thus, we aim to apply this model to the work design. As described in the following sections, human labelers showed excitement on the machine's gradually improving feedback, where labelers' data were trained.

Meanwhile, Rahwan (2018) presented the "society-in-the-loop" model, which reflected upon society's role in the HITL model. Although algorithms make decisions influenced by human users, users are often influenced by stakeholders with conflicting interests and values. Unlike the strong linkage between humans and society reflected in Rahwan's study, we further developed our model – the eHITL model – with humans and communities as separate entities in HITL. The eHITL model emphasizes more the community for a helper of the worker, compared to the previous model. Unlike the "society-in-the-loop" model, our model regards the community and human labeler as separate factors in the loop, respectively, which act differently in the entire process.

In summary, we reviewed recent studies and flows of micro-task designs, tools for better design making, and HITL models considered in human–AI systems. To meet a better labeling environment and algorithmic performance, we want to identify the hidden elements within the labeling process. Moreover, we would like to present a framework for this. Regarding better Our suggested model enhances micro-tasks socially ingrained by employing communities in a crowd work ecosystem and proposes considerations for workers and data diversity for a better tool design. The model also complements the existing HITL model, reflecting fieldwork micro-task design, which further affects human labelers and data qualities. While the previous HITL model had only two stakeholders: the individual worker and the machine, the eHITL model includes an additional group of humans as a stakeholder that positively aids individual crowdworkers. The following section describes how we find the community and machine's feedback as an essential factor to micro-tasks.

**3 STUDY DESIGN**

In this section, we describe the case study and methodology of data collection. We describe how labeling is performed in our problem by analyzing the labelers' behavior and identify any biases in choosing labels.

## 3.1 Method

To find solutions to universal participation and unbiased data quality, we conducted an in-field study to collect log-data, performed in-depth interviews, and were involved in conversations among workers. We participated in and observed a 3-month data labeling project, which was implemented by the Korean Ministry of Science and Information and Communications Technology (ICT). The main task was an image labeling project that aims to make 7 million bounding boxes with 1763 categories of objects from 559 hours of video clips, employed by broadcasting contents. We explored labelers' behaviors by collecting their log data during annotation and conducting surveys and interviews with some participants. The detailed descriptions and methods are as follows.

### 3.1.1 Task and Data

The task that workers performed was an "object/action" labeling in video clips, where they were required to identify at least six labels per clip. The minimum number of labels was arbitrarily determined. The videos contained news programs and entertainment shows, and the videos were split into clips using action units. Thus, the clips' length varied from 2 seconds to 8.3 minutes, and the clips were randomly assigned to the labelers. Labelers started the tasks in August 2020 and finished them in October, and quality assessment continued until November 2020. We observed the behavioral logs from August to October 2020. In early August, all labelers were trained on the basic principles of algorithms and labeling tasks.

In the beginning, there were 2100 subcategories in the labeling, referred to by the official classification documents from multiple sources, such as Statistics Korea and IOC. As they do not apply to fieldwork scenarios, some categories were substituted or removed. Thus, the total number of categories became 1763. The categories' pillars consisted of objects, actions, and situations spread branches into specific labels. For instance, the outdoor facilities were classified as the baseball park, the golf course, and others.

*3.1.2. Participants*

People from various backgrounds participated in the project. Social enterprises participated in this project and hired workers, as presented in Table 1. We called these social enterprises "*communities*," because they acted as community builder in this project. These communities hired and educated people for the task.

Table 1. Description of the communities of workers

| Social enterprises | Common characteristics of workers from each community | Hiring process* | Number of IDs** |
|---|---|---|---|
| A | Elderly people | Personal application | 44 |
| B | Career break women, elderly workers, people with disabilities | Personal application | 11 |
| C | North Korean defectors and an immigrant from the Philippines (all female and under 55) young workers | Recommended by the affiliated organization | 10 |
| D | Young workers | Existing members, part-time job | 5 |
| E | Young workers | Existing members, part-time job | 5 |
| F | People with disabilities, young workers | Existing members | 39 |

*Hiring processes were hold by each enterprise.
**IDs are different from the number of people. IDs were provided to the social enterprises.

Communities comprised of one project leader, one–three educating managers, and multiple temporarily hired workers for the task. Project leaders gathered once or twice a month to participate in committee meetings organized by the project consortium. In these meetings, leaders shared tips on labeling and proposed better user-driven suggestions for labeling by tool developers, such as regulating working hours or using online chat. After the meetings, each leader shared the notes from the committee to their group chatting rooms. Due to the COVID-19 pandemic, most workers primarily used group chats for communication.

Workers frequently talked in group chats and shared tips during the tasks. The project leaders and educating managers constructively participated in communications and often answered questions about labeling criteria. The project manager formulated the guidelines for the tasks, such as the minimum box size of annotations. However, all questions could not be addressed in these guidelines. If all criteria gathered in a book, people would not be able to fully aware of them. Project leaders collected questions and shared them with other communities via chat rooms for discussion, and, then, the project manager updated the guidelines.

Community leaders did not only bridge the gap between the committee and labelers but also confirmed the task outputs. The labeling tool needed human checks avoiding any human errors, and each leader and educating manager verified all outputs. Feedback was provided to the labeling workers almost every day via quick notes in the labeling tool. Frequent errors observed in tasks were also shared to the group chats. During this process, project leaders and educating managers developed their own criteria for labeling, thus suggesting more tactics to both labeling workers and tool developers. Some communities prepared their own manuscripts for the tasks separately.

Therefore, in our observation for 3 months, people substantially relied on the communities they belong to, and communities significantly affected them. With an easy and habituated communication tool – the group chat app in their smart phones, labeling workers could raise questions without any hesitation and readily accept quick answers from their community leaders.

All participants were familiar with using computers and were not visually impaired. The participants' average age was 50 (minimum 20, maximum 68), with a standard deviation of 14.74, and there were 43 female and 32 male participants. According to the employment promotion for the aged act in South Korea, the people aged over 55 are senior citizens. Thus, we used this age criterion. The participants were under contract for 3 months for the project.

Some participants were from different cultural backgrounds. In community C, seven female workers were North Korean defectors, and one female worker was an immigrant from the Philippines. In community B, two young males had different disabilities: one had autism, and the other had a physical disability. Separating them into specific groups was considered for statistical analysis. However, the number of people in specific groups was low. Thus, we note that all participants were distributed in classes based on gender and age for the quantitative study. Then, we applied their specific offers and opinions in the qualitative study described in Section 4.2.

One more note is that community F workers were excluded from our study because their log data were not appropriate for the analysis. People with developmental disabilities in this community could not identify boxes in images effectively. Moreover, they experienced difficulty in choosing six objects independently with subjective decisions, while looking at news images. Furthermore, due to the COVID-19 pandemic, neither face-to-face nor zoom interviews were accepted for the study, due to their weak immune systems and lack of concentration during the remote interview. The details of this issue will be explained in the Limitation sections.

*3.1.3. Procedure*

The participants used laptop computers or desktop PCs offered to each community. Approximately half of the workers worked remotely because of COVID-19, with flexible working hours. The system was accessible at any time. However, some communities, such as Community B, made rules to work for more than 5 hours on weekdays. Owing to some complex issues (e.g., server power down) that arose in the early stage, the participants became accustomed to working between 13:00 and 18:00. In Community B, people gathered once a week and shared tips.

Workers played the clip and browsed for the object to be labeled. To add labels, workers needed to drag the mouse and make a box on the image's object. Then, the proper label for the boxing object was determined. Workers handled videos in which objects move; thus, the objects' coordinates needed to be marked frame by frame. The object tracking program was embedded in a labeling tool; therefore, workers needed to mark the positions of the labels only on the first and last frames. The outputs were stored directly in the database, and we analyzed these collected data.

To explore labels, workers typed the object's name on the category searching box. If no label appeared, they browsed objects in the category list and picked the appropriate one. However, the total number of categories was 1763. Workers mostly memorized which labels were included in the category. They utilized tools such as Microsoft Excel or papers to note frequently appearing objects.

(A) Workers gathering to share tips     (B) Image of the annotation tool

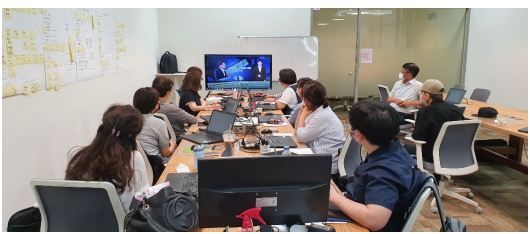 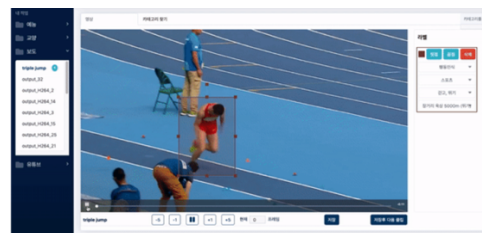

Fig. 2. Workers gathered to share their observations to make better annotations once a week (left). They performed the task with a tool on the right, mainly consisted of a clip menu, video window (labeling area), and category selection section.

*3.2 Analysis*

We did three analyses with this case study to determine the problems and structures for the task design. Workers' log data were fully analyzed by statistical method to identify any distinctive patterns among groups. Workers' interviews, surveys, and communications in the community chat were explored with narrative analysis methodologically. In summary, the entire experimental methods and analysis are described as follows:

(1) Analysis with *data level*: Log data analysis using the chi-square test to measure the differences between demographic groups.
(2) Analysis for *individual level*: Surveys and in-depth interviews of workers.
(3) Analysis for *community level*: A narrative analysis of the workers' community chat for 3 months.

# 4 FINDINGS

With the analyses described above, we present our findings in this section. Our findings can be summarized as follows. First, the labeling pattern showed the differences among groups and communities. Second, workers found choosing labels from several categories complicated. Third, the community positively helped workers in ambiguous situations. We found that the community attempted to contain biased selections of workers. Some workers mistakenly considered that they improved an algorithm's performance, by noticing gradual improvements of the labeling tool they used.

Based on the labelers' log data, different results among groups are introduced in Sections 4.1 and 4.2. We discuss the findings from the in-depth interview concerning workers' individual impressions on the task. Then, we present insights to solve problems from the labelers' communication records, as described in Section 4.3. The last part of Section 4 presents the eHITL model based on the findings discussed above.

*4.1 Data Level: Quantitative Analysis for Labeler's Performance*

This section presents our output that personal biases occur in the labeling output. For the analysis, we first explored the distribution of the number of personal tasks. Second, we split people into groups based on demographic and cultural backgrounds. Finally, we visually checked any distributional differences among the groups.

First, we analyzed the labeler's outputs and log data to verify different patterns among the groups. We used data collected from a labeling tool database, and we determined whether the participants' demographic backgrounds affected the results. The number of tasks were skewed by individuals, as described in Fig. 3 (max: 90,762, min: 874, median: 6,573, average: 10,681). Only five workers did over 20,000 tasks, among the 75 people. Fig. 3 presents the skewed amount of tasks per individual.

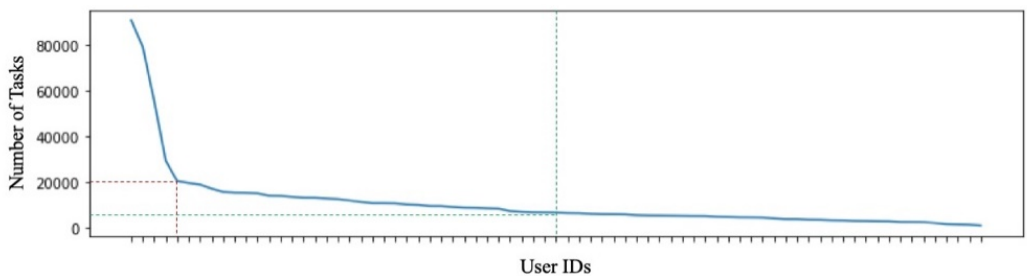

Fig. 3. Skewed number of tasks by personal ID.

Second, we split the IDs into four groups, namely, elderly male, elderly female, young male, and young female, to reveal different patterns among groups. As described previously, people aged over 55 are termed senior citizens in South Korea. The cultural or physical backgrounds were not considered in this category due to the small number of people. We discuss specific requisites from individuals with specific backgrounds in the following section.

For the five outliers shown in Fig. 3, we considered removing the five labelers' data. However, we found that three of them were young females and the others were young males. To quantitatively balance with the other groups, such as elderly groups mainly hired in the labeling task, we omitted the young workers' data from the analysis. Considering that the representativeness could be biased to individual perceptions, we cautiously accessed the younger group's outputs.

We describe the information from the output, such as the total number of labels, variety in labels, and the number of "action" labels, in Table 2. The young female group labeled more actions to video, and both male groups marked a lower variety of labels than the female groups.

Table 2. Analysis of the labeling behavior from the labelers' patterns

|  | Number of people (statistics of age) | Total number of labels | Total number of type of labels | Number of action labels |
| --- | --- | --- | --- | --- |
| Elderly male | 25 (average 60.54, min 55, max 68) | 105,559 | 1,238 | 221 |
| Elderly female | 24 (average 59.29, min 55, max 65) | 158,074 | 1,289 | 230 |
| Young male | 8 (average 30.25, min 25, max 34) | 203,523 | 1,075 | 202 |
| Young female | 19 (average 33.36, min 20, max 54) | 330,790 | 1,295 | 244 |

Third, we visually checked the top 15 labels from each group to confirm the output decision differences. Even with randomly distributed clips, people showed different patterns for labeling objects and actions. In Table 3, the labels in **bold** fonts appeared only in the corresponding groups. The young female group solely showed "the other" in the top 5 categories, indicating that there are no appropriate subcategories for the object. The somatic labels were preferred in all groups, such as head, arm, and upper body. As most of the news data contained the reporter's appearance, labelers often picked features relating to the body.

Table 3. Top 15 labels from each group

| Elderly female    | Elderly male      | Young female        | Young male             |
| ----------------- | ----------------- | ------------------- | ---------------------- |
| Male              | Male              | Male                | Male                   |
| Female            | Female            | Female              | Female                 |
| Microphone        | Upper body        | Necktie             | Necktie                |
| Head              | Chair             | Head                | Head                   |
| Upper body        | Car               | Car                 | Explaining             |
| Chair             | Necktie           | Chair               | Upper body             |
| Explaining        | Explaining        | Upper body          | Microphone             |
| Arm               | Microphone        | Microphone          | Car                    |
| Car               | Outer             | The other           | **Glasses**            |
| Necktie           | Jacket            | Jacket              | Chair                  |
| The other         | Head              | Explaining          | Jacket                 |
| Jacket            | The forties (age) | **Laptop**          | **Answering**          |
| The forties (age) | The other         | Arm                 | The forties (age)      |
| Outer             | **Hat**           | **Announcer**       | **Over the sixties (age)** |
| **Entertainer**   | **Walking**       | **Hosting news program** | **Suit**          |

Finally, we viewed the labeling patterns from the outputs labelled over five times in all groups. Meanwhile, gender-specific objects, action labels, and fashion items showed different labelling patterns among groups. Three researchers selected gender-specific items and cross confirmed the accuracy. For instance, as described in Table 4, elderly people labelled more old-age specific objects, such as traditional crock, Zelkova tree, and plum blossom, while young people chose more situation-related labels such as criminals and cook. Moreover, female workers picked labels more specific to kitchenware.

Table 4. Number of objects expected to be detected by specific genders/ages

|  | Elderly female | Elderly male | Young female | Young male |
|---|---|---|---|---|
| Traditional crock | 50 | 49 | 26 | 3 |
| Plum blossom | 135 | 118 | 0 | 0 |
| Zelkova (tree) | 73 | 77 | 4 | 0 |
| Criminals | 23 | 2 | 175 | 175 |
| Cook (job) | 14 | 8 | 394 | 585 |
| Football player | 31 | 24 | 542 | 209 |
| Dish | 1,522 | 641 | 2,852 | 918 |
| Tray | 81 | 26 | 37 | 1 |
| Cup | 607 | 491 | 1826 | 769 |
| Skillet | 225 | 136 | 387 | 257 |
| Wok | 19 | 8 | 68 | 44 |
| Missile | 36 | 23 | 187 | 136 |
| Fighter (plane) | 19 | 23 | 130 | 79 |

It should be noted that the labeling patterns from the outputs among communities also showed the. To determine which pattern comes first, and whether the differences of gender and age groups preceded that of the community, we also observed more detailed set of labling outputs. For example, we measured gender biases in group A, composed only of elderly people. Results implied that labeling may differ with gender.

Thus, for the answer to **RQ 1.** "How does bias appear when label complexity and various demographic configurations are mixed?", we could answer that, it seems the background of labelers affects various choices. Furthermore, the affiliation to the community also influences the task.

In this study, we focus more on labeling by people with diverse backgrounds. In this field study, labelers dealt with different video clip data in their work. Owing to data difference, labelers may have collected different targets, leading to gaps among groups. Our analysis was based on the assumption that the data distributed to each labeler would be similar. We investigated individual labeling behavior in more in detail, as described in the following section.

*4.2 Individual Level: Surveys to Determine Where They Have Struggled*

To quantify participants' thoughts on the labeling task, we conducted a survey and in-depth interview with 32 participants. After the labeling project has ended, we contacted each community's leader to recruit workers who voluntarily participated in the survey and interview. Thus, the number of answerers decreased, compared to the data analysis. All survey and interview were conducted in corresponding community offices.

In the survey, we proposed five images and asked the participants to label six objects or actions in order. In the in-depth interview, we asked questions about the challenges and difficulties associated with the annotation task. We also asked questions about their technical know-how in performing the task. There were 17 elderly workers (7 females) and 15 young workers (11 females), including two female immigrants from North Korea. The test and in-depth interview took 30 min, and participants received $5 coffee coupons as gratuity.

Table 5. Participants of the survey and in-depth interview.

| Participants | Gender | Age | Participants | Gender | Age |
|---|---|---|---|---|---|
| P1 | Male | 33 | P17 | Male | 39 |
| P2 | Female | 30 | P18 | Female | 27 |
| P3 | Female (From NK) | 26 | P19 | Female | 28 |
| P4 | Female (From NK) | 40 | P20 | Male | 61 |
| P5 | Female | 53 | P21 | Male | 61 |
| P6 | Female | 55 | P22 | Female | 55 |
| P7 | Male | 35 | P23 | Male | 61 |
| P8 | Male | 67 | P24 | Female | 55 |
| P9 | Female | 50 | P25 | Female | 60 |
| P10 | Female | 59 | P26 | Male | 65 |
| P11 | Female | 51 | P27 | Female | 55 |
| P12 | Female | 50 | P28 | Male | 61 |
| P13 | Male | 28 | P29 | Male | 63 |
| P14  | Male   | 60 | P30 | Female | 60 |
| P15  | Female | 27 | P31 | Male   | 69 |
| P16  | Female | 26 | P32 | Male   | 61 |

*4.2.1 Labeling preference based on contextual understanding*

In the survey, participants identified six labels from five images. We first analyzed the labels' frequency filled out and checked the coordinates or sizes of the object affected labelers.

First, people were constantly labeled in clear images. For example, in Table 6, the diversity in labels is apparent in the same position as the person and stable background. In the news program shown on the left, participants picked only 19 labels, while the tv show clip was annotated with 28 labels. There is no audio for the clip, and, therefore, people judged the image on the right in different ways.

Table 6. Six labels marked by labelers for each image.

(A) Image of the news program      (B) Image of the TV show clip

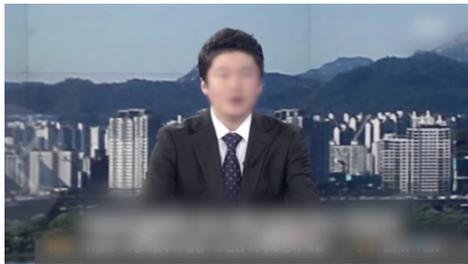 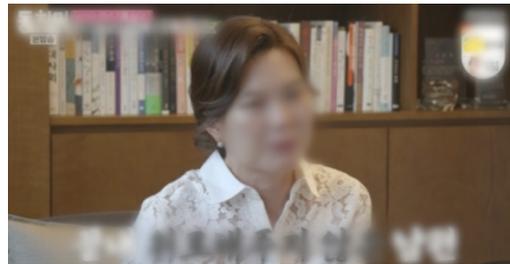

| Top 5 labels **to the left** | Amount | Type | Top 5 labels **to the right** | Amount | Type |
|---|---|---|---|---|---|
| Necktie | 32 | Clothes | Female | 29 | Gender |
| Anchorman | 27 | Job | Blouse | 29 | Clothes |
| Hosting news program | 25 | Action | Book | 28 | Background |
| Male | 24 | Gender | Bookshelf | 25 | Background |

| Jacket | 17 | Clothes | Explaining | 12 | Action |
|---|---|---|---|---|---|

Further, we examined the objects' positions and sizes and how humans' presence influenced workers' decisions. The number of images and survey participants was too small for performing statistical analysis. However, we empirically observed that people do not give importance to the size or position of objects. Instead, they tend to label items with relevance in order. For example, in Fig. 4, workers tend to sequentially label "truck-wheels-number plate" (22 workers) for the left image and "bed-bedding-sofa" (12 workers) for the right. People preferred the image with a person because they could easily annotate their gender, age, job, accessories, and clothes. P5 said, "I prefer to find objects which I could apply multiple labels, such as "woman-age of thirties-upper body-explaining". We can save time to find other objects!" P12 also mentioned that, "I felt lucky when the person with various accessories, such as earrings, a necklace, and glasses appears in the clip! I could quickly fill the minimum number of labels!" However, the person's information, such as job or age, could be subjectively judged, leading to an increase in labels' type.

(A) Image of the news program    (B) Image of the TV show clip

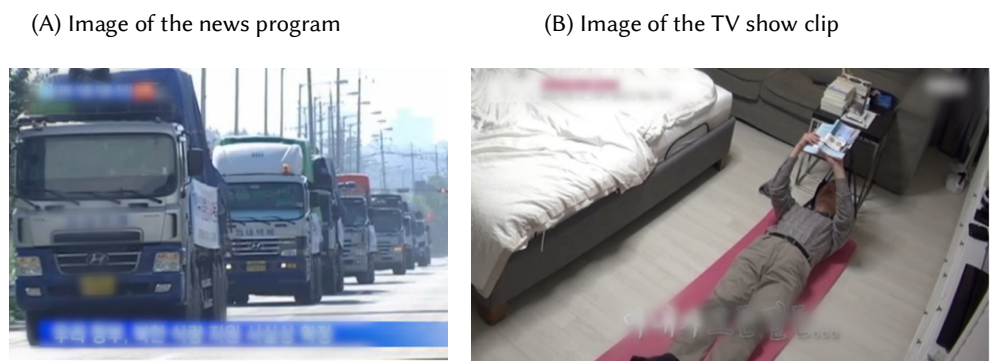

Fig. 4. Images on the left and right received 27 and 23 labels, respectively.

*4.2.2. Personal disagreement on categories.*

After the labeling test, we asked the participants about further challenges in the tasks,

and they asserted disagreements regarding the categories. In particular, the elderly male and young female groups experienced inconvenience in object categorization. In an in-depth interview, the participants expressed that the categories were not descriptive enough to choose proper labels.

They said that the categories were unrealistic. For example, despite that the images have a low resolution to distinguish, there are too many categories listing species of trees, such as pine, oak, juniper, and spruce. However, the variety of vessels is few – there are only ferries in a category; fishing boats or cargo ships were not included. Moreover, there were ambiguities such as safety helmets and gloves being included in the category "fashion items" and rice paddle and trays missing in the "kitchen tools" category. P12 said that "Although we are career-break women, we commonly feel responsible for our everyday tasks. Because of the meticulous nature of women, we cannot bear the loosely structured categories. AI should train more specifically to objects! How are the trees just categorized as a tree in total?"

When such ambiguities arose, participants selected easily discernable objects from their memory. Most elderly workers claimed that they first browsed the categories in the morning, memorized frequently appearing labels, and wrote them down in a notebook next to the monitor. P27 said, "I feel stuffy when I even cannot recall what to search in the category box, while watching the object. I often murmured "ah… what is the name of this" in front of the monitor." P26 also mentioned that "I feel complicated when two appellation was faintly recalled about one action. I even cannot pose an actual example, but, when you get old, you could understand my word." The elderly groups performed annotation step by step, in a calm and orderly manner, compared to the young groups, who seemed to act spontaneously with technical knowledge.

*4.2.3. No hit, no label*

Participants found it difficult to recall the exact category of the labels in the images. To overcome this problem, they browsed the category cell and selected the proper category. Some participants typed one letter in the search box and searched for the relative words. Most young people who are more familiar with technical services used the program's functions, such as relative search.

Moreover, some of the participants were confused during labeling because of different cultural backgrounds. The beret, for instance, which Korean soldiers put on, is quite ambiguous to categorize. In South Korea, it should be categorized as "military goods," but in another perspective (in North Korea), it can simply be interpreted as a hat (fashion item).

Labelers from different cultures also expressed limitations involving language. P4 from North Korea said that "Although the categories were not perfect when facing no proper name for the object, l feel[felt] nervous. For example, I could not search "basket" in a category which was not enrolled in the program, but I thought that was because of my lack of knowledge of South Korean words." P3 said that "In my case, I came to the South in my teens and I went through public education here. So I feel comfortable to Korean standard language including loanwords. However, if defectors or immigrants moved after adolescence might face difficulties to perceive objects in South Korean way." This happened to elderly workers as well. A 63-year-old male worker P29 said that "How do I distinguish the plate, the saucer and the dish? It might sound ridiculous, however, I sometimes feel confused strict meanings." P25 suggested her tip: "I used to type one or two letters to find words in categories. Instead of find words fit into place by force, browsing in list worked much better on me."

*4.2.4. Basic literacy on AI-encouraged labelers*

Interestingly, some participants (7 people) expressed pride in contributing to the improvement of AI. When we asked about the motivation of participation, 13 workers answered because of their interest in AI. Others accidentally found this assignment for a part-time job or were recommended by people around them.

As all labelers in this project were educated about how algorithms train labels beforehand, they all exactly understand what they do and why they label objects. P10 excitedly commented that "I always took account of being helpful to AI's training while annotating labels to data. I hope my work would sufficiently helpful to AI." P23 mentioned that, "I feel so proud that I contributed to deep learning development, while young people avoid to do this simple and boring work. I could tell you that I felt one hundred percent proud of this task." P5 said that "I even feel worry if my "AI kid" trains too well and finally steals human's jobs!"

In conclusion, we could find out any external factors that affect labeling behaviors as follow:

(1) The object which could widely expanded into multiple labels such as human or vehicle could be easily picked from the workers to alleviate their tiredness. This behavioral pattern potentially leads to a biased number of labeled objects in the training dataset.

(2) Ambiguous and disagreeably distributed categories also possibly lead workers to scamp their task. Some workers with better knowledge to AI argue that they should collect only correct labels for a better AI performance.

(3) Optimal functions to complement different personal traits such as cultural background also could affect the task. Without elaborate recommendation tools for label search, people with different mother tongues struggled with what to search. It could build output with biased selections.

*4.3  Community Level: Narrative Analysis from the Labelers' Chat*

To define issues and develop solutions, we analyzed the narratives from the labelers. We explored one community chat where 13 people were involved. The chat room used KakaoTalk, the most popular messaging app in South Korea. Due to privacy concerns, other communities refused to open their chat rooms.

In community B's chat room where we entered, there were two educating managers, one male worker with autism, and ten female workers. With approximately 1100 lines of chats in 3 months, three researchers classified and coded the text into several topics. They were mainly complaining about the labeling tool's errors and physical fatigue. In particular, labelers shared annotation tips spontaneously. When they encountered something new or intriguing in their working tool, they shared the news without hesitation. They acted similar to one-team even though the task was definitely an individual work. We focused on the team-playing phenomenon in the narratives.

*4.3.1. Quick and active comments preventing labelers from task abandonment*

In the chat room narrative, in addition to the educating managers and leaders, the other labelers attempted to solve problems and made quick and proper decisions. If uncertainty occurred for individuals, the community determined the criteria for classification immediately. For example, one of the labelers asked, "The reporter wears a firefighter's uniform in the news program. How do I label it? Is he a reporter, or a firefighter?" Other labelers comments with their ideas, such as "In my opinion, our task is for training AI. So, even we label differently, AI would classify reporter and firefighter with other contextual information in the clip, due to the massive amount of labeled data (labeler, female, 55)." Then, a manager gathered opinions, asked the project manager, and quickly replied "If there is no microphone held in hand, it should be annotated as a firefighter." Workers mutually encouraged one another.

Moreover, labelers devised their own principles in confusing moments. They used a metaphor in defining some rules in labeling. As the community mostly contained elderly females with kids, when somebody abandoned tough tasks, they always recalled that "Suppose that we are explaining six-year-old kid about the object or situation in the image. Our guidance will improve kid's perception to the world."

*4.3.2. Community may beat ambiguity*

Moreover, there were ambiguous criteria present in the categorization. For labelers who wanted to annotate objectively, categories that were difficult to select were a major issue. In narratives, ambiguity is revealed as an inducement for subjective decisions because labelers easily became tired of agonizing. In addition, the insufficient labeling tool and positioning opposite sides of the community are hurdles for annotation.

Communities mediated the gap between individuals' subjective decisions and detailed annotations. In the possibility of subjective decision to lean over specific labels, communities intervened in the tasks and arranged solutions to maintain objectivity. For example, when the labels leaned too much toward somatic objects, manager (male, 34) mentioned that "I observed "standing" label moments ago, and this kind of "action" label could be helpful on making up the number of labels."

Moreover, the numbers that should be marked per clip seemed to work effectively on inducing various annotations. More than six labels should be given to one clip; therefore, workers needed to identify multiple objects and actions from the images. When labelers complained about the minimal count of label, community managers helped with them in finding more labels using video conferencing tools such as Zoom.

*4.3.3. Feeling interactive between the AI and I*

Notably, labelers treated the labeling tool and algorithms as kids. As their principles lay in "six-year-old kid training," workers noticed gradual changes in labeling tools as an algorithmic improvement based on their effort. One labeler (female, 59) said that "Once the labeling box moved naturally frame by frame, I perceived that the tool got better by training our output. Then, I realized that our performances could be used [for] object tracking or detecting systems." Another labeler (female, 53) said "By observing improvement of the tool, I am pretty certain that AI train quite well our labels!"

As they worked on the frontline of the task, labelers physically perceived the tool's mechanical improvements as specifically enhancing user (labeler) experience, such as increased server speed and reduced lag between image frames. However, the tool's update was not related to the data collected by the labelers. The program was updated for better labeling and not affected by any algorithmic sources.

We interpreted that even a worker's misunderstanding could act as a motivational effect. People felt their fruitful outcomes in their own way; thus, they showed more active and positive participation in the task.

Therefore, we noticed how do workers select labels from a large number of categories as follows. Individual workers memorize frequently appearing labels with their external devices such as excel files or notebooks, and positively share their tips with their colleagues. By interchanging ideas in the community, workers become aware of more diverse items in the image. To the **RQ 2**, "Are there any invisible influencing factors in the current complex labeling process?," we could address the invisible stakeholders such as community as an answer.

## 5 Our Design Suggestion: Extended Human-in-the-Loop Model

Based on our findings, we designed the eHITL model for task design, as described in Fig. 1. In general, the HITL model only contains interactive feedback between the human and the machine. We extended the loop to the community beyond the labeler. This model includes a wide range of stakeholders involved in labeling tasks. Individuals as well as developers and administrators who keep their projects stable participate in micro-tasking.

### *5.1. Community as a component of the loop.*

Comparing with Rahwan's work (2018), we did not contain individual workers into society range. By separating worker and community as different actants inside the loop, we could more easily understand perspective, respectively. As Tubaro et al. (2020) asserted, workers in the community, such as project leader and education manager, could take multiple invisible roles like "the trainer, verifier, and the imitator." Our in-depth interviews described in Section 4.2 also showed that the need for communities for the individual workers. In addition, based on our findings, the community could prevent potential bias or high subjectivity from the individuals' detailed annotation tasks. Thus, community work is meaningful in HITL models in labeler and machine interaction.

### *5.2. Machine's feedback could act as a reward*

In addition, we should consider the machine's feedback. In the interview, the participants mentioned that the machine's apparent progression was quite impressive, validating their self-esteem on the task performance. In particular, some workers said their motivation for task participation was curiosity about AI. They wanted to know how they could contribute to AI's development. From their answers from the interview, we observed that they perceived sensitively to the changes of the tool. Even though it is not an exact improvement that emerged from the algorithm's better performance, some gradual mechanical changes can lead to valuable emotions in workers.

### 5.3. Edges in the loop

The eHITL model also includes interactions between actants, described as edges between nodes. When labeler asks certainty of their decisions to label something to the community, this action could solely contribute to mitigating workers' subjective biases. More and more proper feedback from the community could reduce participation difficulties and gain universality to the public task. While obtaining better training data, the machine could perform better outputs. With this virtuous circle, skilled human resources for proper labeling also could be created.

## 6 DISCUSSION

In this paper, we presented the eHITL model for micro-task design. For identifying the challenges of a detailed labeling task, we conducted an in-field study. Unlike simple annotation tasks reported in previous studies, we uncover a unique labeling design model by workers with diverse backgrounds, established based on the connection with the individual labeler, community, and machine. In this section, we discuss how our findings and model suggest key directions for further research on the micro-work ecosystem.

### 6.1 Presence of the Loop

Our interviews and narrative analysis revealed that the individual labeler did not solely perform their task. They are affected by the design of an annotation tool, as previous studies presented. They may also encountered difficulty when facing ambiguous criteria of labeling categorization. With unexpected feedback from the machine, they could feel embarrassed or rewarded. Stakeholders in the micro-task did not behave separately. They positively interacted with others.

We found the loop among actants in the labeling project, which seemed to not generally handled in previous studies. Communities, being regarded as recruitment sources, took an important part in management to labelers and labeling practices. They triggered workers do work better and sustain their tasks. Meanwhile, the machine's real-time feedback took charge of improving the labeler's motivation. They created the loop for a better task design. Looking into the stakeholders' reestablished roles could implicate broader insights into both micro-task studies and labeling projects.

*6.1.1. Interactive nudge to improve worker's pride*

Nudging workers to perform proper labeling showed a better mood in the community. Because it is only penny-earning work, labelers could easily quit the task if they do not find it attractive. For communities, however, stacking skilled labelers and making a good labeler pool is essential. To sustain tasks, communities nudge to labelers to make a better performance, instead of giving orders. They offer further educations, such as "what labels do for AI's operation" or "the value of the gig economy" to make workers develop their digital literacy.

Interestingly, by knowing more about labeling tasks' value, workers showed pride in their contributions to the near future improved by AI development and regarded themselves as a leading role in ML's better performance. With an increment of digital literacy, workers understood and even misunderstood the annotation tool's updates as gradual improvement, powered by labeler's output. These understandings/misunderstandings encouraged workers to perform their work seriously.

In terms of decency, the workers' pride should be strongly considered as well. Some workers felt reluctant about participating in jobs that pay less and seem to be trivial because they tend to be conscious of others' judgment. Therefore, communities or task organizers need to offer sufficient information about how their tasks contribute to future generations. It is also crucial that decisions affected by various demographic backgrounds be identified to prevent algorithmic bias.

*6.1.2. Fair participation vs. individual decision-making*

As shown in Table 4, workers from different demographic groups showed various results in labeling tasks. The result offers the possibility that gathering more diverse types of workers could bring more various labels. Even the community controls the labeler's subjective decision-making, it could not be easy to standardize the whole tasks.

Community's guidelines and controls prevent workers from the absentminded working and uncertainty on selecting labels. Not dropping out any workers who lack both physical and mental energy to sustain the work, communities also contribute to guarantee fair participation by encouraging and supporting people, leading to expanding the participator's pool.

Individual decision-making could be affected by their communities naturally. As described in Section 4.3, people gathered and shared tips for labeling. Also, workers behaved to refer to or even imitate others' decision-making to ensure the quality of their outputs. As Van Berkel et al. (2019) asserted that decision making in social group with diverse co-labeling pools achieve higher co-agreement, our findings also show that the mediation in communities could occur unified outputs, logically. Further study should consider the gaps in the results among the communities.

*6.2 Design Implications*

Our findings revealed the essential factor that task designers could take to enhance the participation of people with diverse backgrounds and mitigate potential biases in detailed labeling tasks. Thinking of loop among the community-labeler-machine could benefit algorithm's performance by expanding the pool of workers and guaranteeing qualities of the labels.

*6.2.1. Considering [invisible] stakeholders*

We illustrated the structured relationships in the loop in the annotation task with specifically categorized labels in Section 3.4. We could bring up invisible stakeholders – the community and the machine. By interpreting the interactions between actants, we could further explore the main challenges for the task.

The eHITL model includes main stakeholders for detailed labeling tasks. However, an orderer who organized the project is not in the loop. It is clear that the AI designer or developer controls labelers with the guidelines. In our study, labelers experienced difficulty in understanding their orders. Therefore, communities interpreted manuals easily understood for everyone. As an AI designer, we should also look outside the box and leave the loop work smoothly.

*6.2.2. Needs for annotation tools differed for diverse cultural backgrounds*

Given these findings, designers should consider that labelers with different backgrounds required various requirements for the labeling tool. For example, the North Korean defector group asked for typo-correction options. With an apparent orthographic difference, especially in the 1940s such as [j+j], issues occurred more often in loanword orthography. Similar spelling issues were also prevalent in elderly workers. Owing to several Korean orthography reforms, some people who did not learn new rules after completing their formal education tend to write wrong spellings habitually.

For the people with developmental disabilities, on the other hand, they need object recommendations. We observed that they find labeling an image difficult. Thus, the managers of the community simplified their work. For example, a manager instructed workers to annotate "a head" or "a car" in multiple images. For workers who could not spontaneously select labels, object recommendation for annotating should be strongly considered, as well.

*6.3 Guidelines for the labeling project*

Here, we suggest a guideline for the labeling project, which contains a large number of various workers and a massive amount of categories. First, organize subcommunities directly communicate with labelers. As we described above, communities could help marginalized people and drive more varied decisions when performing the tasks. Second, carry out the education for workers to increase AI literacy. Better knowledge could consequently bring them better motivation. Third, show mechanical changes or improvements to workers continuously. People feel rewarded by physically noticing their outputs playing a role in AI's development. Fourth, refer to the eHITL system for a quick reflection on on-site feedback. The sooner improvement of labeling tool could bring better performance to algorithms. By incorporating stakeholders in the project, constant feedback in the loop could lead to flawless output. External devices such as official group chat apps could be used properly.

*6.4 Limitations and future work*

There were some challenges involved in our approach. At an early stage, because of an unstabilized labeling program, workers felt confused and exhibited gaps in their outputs, depending on their learning capabilities. Some issues arose, making it challenging to explore the sequential log data of workers' behavioral changes, so we could not statistically measure the influence of the program's updates on labeler's output. Thus, we relied on worker's wording instead in that they felt rewarded by tool improvement.

Either, unlike other communities, the log data from people with developmental disabilities could not be analyzed. As described above, most of them could not use a single ID because of the task's complexity. Workers used the same IDs in turn; thus, the IDs could not be considered an individual's use. Further, because of the COVID-19 pandemic, we could not directly interview the participants. In extended periods of non-face-to-face work for the socially disadvantaged, we should develop alternative methods for people who are not physically available to use such IT tools.

Due to privacy concerns, the government (the project provider) strongly regulated sharing of private information, such as education and social-economic status. Also, we could not access limited individual factors. We should try various features affecting labelers' behaviors in a controlled environment in future work.

This paper proposed a task design model; however, we could not examine the statistical comparison. This is because our case was based on a fieldwork project, which cannot be controlled for the study. Hence, we conducted in-depth research using qualitative methods to identify the requirements for the in-situ task. In future work, we will apply our model to a structured experiment and examine its performance.

In a future study, we will also explore if this model fits other crowdsourcing tasks such as voice recording or summarizing texts. Our study possibly showed exceptional results because of a particular cultural sphere. However, we expect this local case's findings could be generalized in global circumstances, especially in the countries where micro-works are dawning. We hope that our design attributes will ensure easy access to social labeling works and guarantee better data quality.

## 7  CONCLUSIONS

This study presents the eHITL model for socially embedded work for detailed data annotation. The model accounts for the labeler's cultural/demographic diversity and minimizes the hurdle in participation by adopting communities in the system. Further, the machine's real-time feedback, which validates the participants' contributions to the output, has been emphasized. Our design will help increasing the diversity in micro-tasks and enhancing ML's performance trained by using the human-made dataset. Furthermore, it will contribute to the data labeling ecosystem, preventing algorithmic bias induced by a lean over the specific background.

**ACKNOWLEDGMENTS**

We would like to thank the consortium members and workers who helped us in providing the material, as well as spiritually. The consortium included a project management team, a labeling tool development team, researchers, and six social enterprises. We also thank the editors and anonymous reviewers of *International Journal of Human-Computer Interaction* for their helpful comments.

**Disclosure statement**

No potential conflict of interest was reported by the author(s).

**Funding**

This was not a funded study.

# Biographies

**Jaeyoun You**

Jaeyoun You is a Ph.D. candidate in the Department of Intelligence and Information at Seoul National University, Seoul, South Korea. Her research interests include algorithmic bias, image analysis, and human-AI collaboration.

**Daemin Park**

Daemin Park is an Assistant Professor in the School of Media & Communication at Sunmoon University(2021-present). He was a Chief Technology Officer at Korea Data Exchange(2019-2020). He also worked as a Senior Research Fellow at Korea Press Foundation(2014-2019) and a staff reporter at Maeil Business Newspaper(2006-2012).

**Joo-yeong Song**

Joo-yeong Song is a data scientist at Obzen, Seoul, South Korea. She received her M.S. from the Department of Intelligence and Information at Seoul National University. Her research interests include recommendation systems and social media data.

**Bongwon Suh**

Bongwon Suh is a Professor in the Department of Intelligence and Information at Seoul National University, Seoul, South Korea. His research interests include human-computer interaction, social computing, and data analysis.